\documentclass[aps]{revtex4}%
\usepackage{amsfonts}
\usepackage{amsmath}
\usepackage{amssymb}
\usepackage{graphicx}%
\providecommand{\U}[1]{\protect\rule{.1in}{.1in}}
\providecommand{\U}[1]{\protect\rule{.1in}{.1in}}

\newcommand{\be}{\begin{equation}}
\newcommand{\ee}{\end{equation}}
\newcommand{\bea}{\begin{eqnarray}}
\newcommand{\eea}{\end{eqnarray}}

\begin{document}
\title{A Decomposition of Separable Werner States}
\author{R.G. Unanyan$^{1,2}$, H. Kampermann$^{1}$, and D. Bru\ss $^{1}$}
\author{}
\affiliation{$^{1}$Institut f\"{u}r Theoretische Physik III, Heinrich-Heine-Universit\"{a}t
D\"{u}sseldorf, D-40225 D\"{u}sseldorf, Germany}
\affiliation{$^{2}$Institute for Physical Research, Armenian National Academy of Sciences,
378410 Ashtarak, Armenia}

\begin{abstract}
We derive an integral convex combination of product states for a range of
separable Werner states. Our method consists of expanding the sought-after
local density operators in terms of Wigner operators. For dimension $d=2$, our
decomposition holds for the whole separable range of Werner states, while for
$d>2$ it is valid for a subset of separable Werner states. We illustrate the
general method with the explicit examples $d=2$ and $d=3$.

\end{abstract}
\date{08.03.2007}
\startpage{1}
\maketitle

\section{\bigskip Introduction}

Composite quantum systems are among the foci of quantum information theory.
Bipartite quantum states, i.e. states of quantum systems that consist of two
subsystems A and B, can be classified according to their property of being
entangled or separable. In 1989, R. Werner proposed a physically meaningful
definition of separability of a quantum state \cite{Werner}, namely
\begin{equation}
\rho_{sep}=\sum_{i}p_{i}\rho_{A}^{i}\otimes\rho_{B}^{i}\ ,
\end{equation}
where $p_{i}$ are probabilities, i.e. $p_{i}\geq0$ and $\sum_{i}p_{i}=1$. In
the integral version of this convex combination of product states, the
discrete index $i$ is replaced by some continuous variable $\lambda$, such
that the integral product decomposition reads $\rho_{sep}=\int d\lambda
\,p(\lambda)\rho_{A}(\lambda)\otimes\rho_{B}(\lambda)$, where $p(\lambda
)\geq0$ and $\int d\lambda\,p(\lambda)=1$. Given a separable state, its
product decomposition is not unique, and in general, it is a hard task to find
such a decomposition. If one does not know whether a given state is separable
or entangled, it is difficult to prove whether a separable decomposition does
or does not exist. This is the origin of the entanglement vs. separability
problem \cite{primer,db}.

In his seminal paper \cite{Werner}, Werner also introduced a certain family of
states that is nowadays referred to as Werner states. Due to their symmetry
properties, they play an important role in several contexts, e.g. in
entanglement purification \cite{puri}, entanglement properties for states with
white noise, and the possible existence of bound entangled states with
non-positive partial transpose \cite{npt}.

For qubits, a separable decomposition of Werner states has been found in
\cite{wootters}. A different decomposition was recently derived in
\cite{azuma}, again for 2-dimensional subsystems. Here, we find a separable
decomposition of Werner states in arbitrary dimensions. For dimension $d=2$,
our decomposition is different from both decompositions in \cite{wootters} and
\cite{azuma}.

\section{Werner States}

Werner states \cite{Werner}, in the following denoted as $\rho_{W}$, are a
class of mixed states for bipartite quantum systems (where each of the two
subsystems has dimension $d$), which are invariant under the transformations
$U\otimes U$, for any unitary $U$, i.e. $\rho_{W}=(U\otimes U)\rho
_{W}(U^{\dagger}\otimes U^{\dagger})$. The family of Werner states,
characterized by one parameter $f$, is (in the original notation) given by the
density operator
\begin{equation}
\rho_{W}=\frac{1}{d^{3}-d}\left(  \left(  d-f\right)  {{1\!\!1}}+\left(
df-1\right)  \mathbf{V}\right)  \ , \label{WernerState}%
\end{equation}
which acts on the $d\times d$-dimensional Hilbert space $H_{A}\otimes H_{B}$
that is spanned by the state vectors of subsystems $A$ and $B$. Here,
${1\!\!1}$ is the identity operator on $H_{A}\otimes H_{B}$, and by
$\mathbf{V}$ we denote the swap operator, which acts as $\mathbf{V}%
|\,\phi\rangle_{A}\otimes|\,\psi\rangle_{B}=|\,\psi\rangle_{A}\otimes
|\,\phi\rangle_{B}$. Positivity of $\rho_{W}$ implies for the parameter
$f=tr(\rho_{W}\mathbf{V})$ that $-1\leq f\leq1$. The Werner state $\rho_{W}$
is separable, i.e. classically correlated, iff $0\leq f\leq1$. In the
following, we will regard the Hilbert space of a subsystem as the state space
of a particle with spin $j$, where $d=2j+1$. The basis states are denoted by
$\left\vert jm\right\rangle $, where $m=j,j-1,...-j$. It is known
\cite{Biedenharn} that $\mathbf{V}$ has the following form in terms of the
Wigner operators $T_{q}^{k}$ \cite{Biedenharn}:%
\begin{equation}
\mathbf{V}=%
{\displaystyle\sum\limits_{k=0}^{2j}}
{\displaystyle\sum\limits_{q=-k}^{k}}
\left(  -1\right)  ^{q}T_{q}^{k}\otimes T_{-q}^{k}\ , \label{permutation}%
\end{equation}
where $T_{q}^{k}$ denotes the $\left(  2j+1\right)  \times\left(  2j+1\right)
$ matrix with the elements \ given by \cite{Biedenharn}
\begin{equation}
\left\langle jm_{1}\right\vert T_{q}^{k}\left\vert jm_{2}\right\rangle
=\left(  \frac{2k+1}{2j+1}\right)  ^{1/2}C_{m_{1}qm_{2}}^{j\,k\,j}.
\label{tensoroperators}%
\end{equation}
The\ coefficients $C_{m_{1}qm_{2}}^{j\,k\,j}$ are Clebsch-Gordan coefficients.
We use the notation $|\,JMj_{1}j_{2}\rangle=\sum_{m_{1},m_{2}}C_{m_{1}m_{2}%
M}^{j_{1}\,j_{2}\,J}|\,j_{1}m_{1}\rangle\otimes|\,j_{2}m_{2}\rangle$, where
$J$ is the total angular momentum, $M$ its third component, $j_{i}$ is the
angular momentum of particle $i$, and $m_{i}$ its third component, with
$m_{1}+m_{2}=M$. Note that the $\left(  2j+1\right)  ^{2}$ operators
$T_{q}^{k}$ generate the group $U\left(  2j+1\right)  $. The matrices
$T_{q}^{k}$ obey the orthogonality relations
\begin{equation}
tr\left(  T_{q_{1}}^{k_{1}}\cdot\left(  T_{q_{2}}^{k_{2}}\right)  ^{\dagger
}\right)  =\delta_{k_{1}k_{2}}\delta_{q_{1}q_{2}} \label{orthogonality}%
\end{equation}
and are traceless\ for $k\neq0$ \cite{Biedenharn}. For $k=0$ we have
\begin{equation}
T_{0}^{0}=\left(  2j+1\right)  ^{-1/2}{1\!\!1}. \label{identity}%
\end{equation}
Using Eq. (\ref{permutation}), we can rewrite the Werner state
(\ref{WernerState}) in terms of the Wigner operators:%
\begin{equation}
\rho_{W}=\frac{1}{d^{3}-d}\left(  \left(  d-f\right)  {{1\!\!1}}+\left(
df-1\right)
{\displaystyle\sum\limits_{k=0}^{2j}}
{\displaystyle\sum\limits_{q=-k}^{k}}
\left(  -1\right)  ^{q}T_{q}^{k}\otimes T_{-q}^{k}\right)  \ .
\label{Werener2}%
\end{equation}

\section{Decomposition of separable Werner states}

In general, a separable state can be written in the integral decomposition
\cite{Werner}
\begin{equation}
\rho_{sep}=\int d\lambda\, p\left(  \lambda\right)  \left[  \rho_{A}\left(
\lambda\right)  \otimes\rho_{B}\left(  \lambda\right)  \right]  ,
\label{decomposition}%
\end{equation}
where
\[
p\left(  \lambda\right)  \geq0,\text{ \ \ }\int d\lambda\, p\left(
\lambda\right)  =1\ ,
\]
and $\rho_{A}\left(  \lambda\right)  \geq0$, $\rho_{B}\left(  \lambda\right)
\geq0$ represent density operators of subsystems $A$ and $B$. Here, $\lambda$
has to be understood as a symbol for one or more continuous variables.

We can express an arbitrary density operator $\rho\left(  \lambda\right)  $ of
a spin-$j$ particle via the operator decomposition
\[
\rho\left(  \lambda\right)  =%
{\displaystyle\sum\limits_{k=0}^{2j}}
{\displaystyle\sum\limits_{q=-k}^{k}}
y_{kq}\left(  \lambda\right)  T_{q}^{k}\ ,
\]
where the expansion coefficients $y_{kq}$ are given by
\begin{equation}
y_{kq}(\lambda) = tr \left(  \rho(\lambda) T_{q}^{k\, \dagger} \right)  \ ,
\end{equation}
due to the orthogonality relation (\ref{orthogonality}).

Using this decomposition for both $\rho_{A}(\lambda)$ and $\rho_{B}(\lambda)$
we can rewrite a separable Werner state (\ref{Werener2}) as%
\begin{equation}
\rho_{W}=\int d\lambda\,p\left(  \lambda\right)  \left[
{\displaystyle\sum\limits_{k_{1}=0}^{2j}}
{\displaystyle\sum\limits_{k_{2}=0}^{2j}}
{\displaystyle\sum\limits_{q_{1}=-k_{1}}^{k_{1}}}
{\displaystyle\sum\limits_{q_{2}=-k_{2}}^{k_{2}}}
y_{k_{1}q_{1}}\left(  \lambda\right)  y_{k_{2}q_{2}}\left(  \lambda\right)
T_{q_{1}}^{k_{1}}\otimes T_{q_{2}}^{k_{2}}\right]  . \label{integralform}%
\end{equation}
A comparison of (\ref{Werener2}) with (\ref{integralform}) leads to the
following condition on $y_{k_{1}q_{1}}\left(  \lambda\right)  $ and
$y_{k_{2}q_{2}}\left(  \lambda\right)  $:
\begin{equation}
\int d\lambda\,p\left(  \lambda\right)  y_{k_{1}q_{1}}\left(  \lambda\right)
y_{k_{2}q_{2}}\left(  \lambda\right)  \sim\delta_{k_{1},k_{2}}\delta
_{q_{1},-q_{2}}. \label{spheric}%
\end{equation}
We can conclude that $y_{k_{i}q_{i}}\left(  \lambda\right)  $ with $i=1,2$ are
orthogonal, with a weight function $p\left(  \lambda\right)  $. Remember that
the Werner state on the left-hand side of Eq. (\ref{integralform}) depends on
the parameter $f$, which is not explicitly written here. Thus, the expansion
coefficients $y_{k_{1}q_{1}}\left(  \lambda\right)  $ and $y_{k_{2}q_{2}%
}\left(  \lambda\right)  $ and the weight function $p\left(  \lambda\right)  $
will in general be functions of $f$ as well.

It is possible to satisfy the orthogonality condition in Eq. (\ref{spheric})
in the following way: let us consider $\lambda$ as a symbol for two
parameters. By defining $\lambda=\{\theta,\varphi\}$ with $0\leq\theta\leq\pi$
and $0\leq\varphi\leq2\pi$ and $p\left(  \theta,\varphi\right)  =1/4\pi$
\cite{azuma} the relation (\ref{spheric}) takes the form%
\begin{equation}%
{\displaystyle\int\limits_{0}^{\pi}}
d\theta%
{\displaystyle\int\limits_{0}^{2\pi}}
d\varphi\sin\theta y_{k_{1}q_{1}}\left(  \theta,\varphi\right)  y_{k_{2}q_{2}%
}\left(  \theta,\varphi\right)  \sim\delta_{k_{1},k_{2}}\delta_{q_{1},-q_{2}}.
\label{orthogonality2}%
\end{equation}
Thus we can use for $y_{kq}\left(  \theta,\varphi\right)  $ the spherical
harmonics $Y_{q}^{k}\left(  \theta,\varphi\right)  $. One can readily check
that the decomposition (\ref{integralform}) can be rewritten as%
\begin{equation}
\rho_{W}=\frac{1}{4\pi}%
{\displaystyle\int\limits_{0}^{\pi}}
d\theta%
{\displaystyle\int\limits_{0}^{2\pi}}
d\varphi\, \sin\theta\left[  \rho_{A}\left(  \theta,\varphi\right)
\otimes\rho_{B}\left(  \theta,\varphi\right)  \right]  ,
\label{decomposition0}%
\end{equation}
where
\begin{equation}
\rho_{A}\left(  \theta,\varphi\right)  =\frac{1}{2j+1}{1\!\!1}+\frac
{1}{\left(  2j+1\right)  }\frac{\left(  2j+1\right)  f-1}{\left(  2j+1\right)
^{2}-1}%
{\displaystyle\sum\limits_{k=1}^{2j}}
{\displaystyle\sum\limits_{q=-k}^{k}}
\eta_{kq}^{-1}Y_{q}^{k}\left(  \theta,\varphi\right)  T_{q}^{k}, \label{first}%
\end{equation}

\begin{equation}
\rho_{B}\left(  \theta,\varphi\right)  =\frac{1}{2j+1}{1\!\!1}+%
{\displaystyle\sum\limits_{k=1}^{2j}}
{\displaystyle\sum\limits_{q=-k}^{k}}
\eta_{kq}Y_{q}^{k}\left(  \theta,\varphi\right)  T_{q}^{k}, \label{second}%
\end{equation}
with $\eta_{kq}$ being arbitrary parameters. We note that hermicity of
$\rho_{A,B}\left(  \theta,\varphi\right)  $ implies that $\eta_{kq}^{\ast
}=\eta_{k,\text{ }-q}.$ We have used the properties of the spherical harmonics
$Y_{q}^{k}\left(  \theta,\varphi\right)  ,$ namely Eqs.(\ref{orthogonality2})
and $Y_{q}^{k\ast}\left(  \theta,\varphi\right)  =\left(  -1\right)
^{q}Y_{-q}^{k}\left(  \theta,\varphi\right)  $. The above combination of
product states for $\rho_{W}$
still contains the freedom in choosing the parameters $\eta_{kq}$. Note,
however, that at this point we have not yet shown positivity of the local
operators, and the notation $\rho_{A}$ and $\rho_{B}$ is merely suggestive.
The reader will have noticed that the prefactors of the sums in expressions
(\ref{first}) and (\ref{second}) are not identical. We have chosen this
asymmetric decomposition on purpose, as will be explained below. The
decomposition (\ref{decomposition0}) could have equally well been formulated
in a symmetric way by defining
\begin{equation}
\rho_{A}^{\left(  s\right)  }\left(  \theta,\varphi\right)  =\frac{1}%
{2j+1}{1\!\!1}+\sqrt{\frac{1}{\left(  2j+1\right)  }\frac{\left(  2j+1\right)
f-1}{\left(  2j+1\right)  ^{2}-1}}%
{\displaystyle\sum\limits_{k=1}^{2j}}
{\displaystyle\sum\limits_{q=-k}^{k}}
{\eta^{\prime}}_{kq}^{-1}Y_{q}^{k}\left(  \theta,\varphi\right)  T_{q}^{k},
\label{firstsymm}%
\end{equation}%
\begin{equation}
\rho_{B}^{\left(  s\right)  }\left(  \theta,\varphi\right)  =\frac{1}%
{2j+1}{1\!\!1}+\sqrt{\frac{1}{\left(  2j+1\right)  }\frac{\left(  2j+1\right)
f-1}{\left(  2j+1\right)  ^{2}-1}}%
{\displaystyle\sum\limits_{k=1}^{2j}}
{\displaystyle\sum\limits_{q=-k}^{k}}
\eta_{kq}^{\prime}Y_{q}^{k}\left(  \theta,\varphi\right)  T_{q}^{k}.
\label{secondsymm}%
\end{equation}

In this version $\rho_{A}^{\left(  s\right)  }\left(  \theta,\varphi\right)  $
and $\rho_{B}^{\left(  s\right)  }\left(  \theta,\varphi\right)  $ have a nice
symmetric form, however their positivity conditions are quite difficult to
analyze.
It is obvious that the separability property of the Werner state does not
depend on the form of the local density operators, therefore we choose to use
the local operators (\ref{first}) and (\ref{second}), because they allow to
determine the positivity constraints in an easier way.

If all eigenvalues of the local operators (\ref{first}) and (\ref{second}) are
positive, we found a valid separable decomposition of the Werner state
$\rho_{W}$ in Eq. (\ref{decomposition0}). The calculation of the eigenvalues
of the local density operators for $j\geq1$ is a difficult task. We now
simplify the analysis by choosing $\eta_{kq}=\eta_{k}$. Since
\begin{equation}
Y_{q}^{k}\left(  \theta,\varphi\right)  =\sqrt{\frac{2k+1}{4\pi}}%
D_{q0}^{k\ \ast}\left(  \theta,\varphi\right)  \ , \label{spheric2}%
\end{equation}
where $D_{qm}^{k}\left(  \theta,\varphi\right)  $ is the Wigner rotation
matrix, defined via the transformation of $T_{q}^{k}$ as a tensor
\cite{Biedenharn}, i.e.
\begin{equation}
U:T_{m}^{k}\rightarrow UT_{m}^{k}U^{\dagger}=%
{\displaystyle\sum\limits_{q=-k}^{k}}
D_{qm}^{k}\left(  \theta,\varphi\right)  T_{q}^{k}\ . \label{tensor}%
\end{equation}
Inserting Eq. (\ref{spheric2}) into Eqs. (\ref{first}) and (\ref{second}),
using Eq. (\ref{tensor}) and the fact that $D_{qm}^{k\ \ast}\left(
\theta,\varphi\right)  =D_{qm}^{k}\left(  \theta,-\varphi\right)  $, we can
rewrite the local density matrices in the form%

\begin{equation}
\rho_{A}\left(  \theta,\varphi\right)  =U_{A}\left(  \theta,-\varphi\right)
\left(  \frac{1}{2j+1}{1\!\!1}+\frac{1}{\left(  2j+1\right)  }\frac{\left(
2j+1\right)  f-1}{\left(  2j+1\right)  ^{2}-1}%
{\displaystyle\sum\limits_{k=1}^{2j}}
\eta_{k}^{-1}\sqrt{2k+1}T_{0}^{k}\right)  U_{A}^{\dagger}\left(
\theta,-\varphi\right)  ,
\end{equation}%
\begin{equation}
\rho_{B}\left(  \theta,\varphi\right)  =U_{B}\left(  \theta,-\varphi\right)
\left(  \frac{1}{2j+1}{1\!\!1}+%
{\displaystyle\sum\limits_{k=1}^{2j}}
\eta_{k}\sqrt{2k+1}T_{0}^{k}\right)  U_{B}^{\dagger}\left(  \theta
,-\varphi\right)  .
\end{equation}
Here, $U\left(  \theta,\varphi\right)  $ denotes the unitary irreducible
representation of the $SO\left(  3\right)  $ group on the state space spanned
by $\left\vert jm\right\rangle $, with $m=j,j-1,...,-j$, and is defined as
\[
U\left(  \theta,\varphi\right)  =\exp\left(  -i\varphi J_{z}\right)  \cdot
\exp\left(  -i\theta J_{y}\right)  \ .
\]
Note that in the unitary operator we have already omitted the third Euler
angle $\gamma$, by dropping the factor $\exp\left(  i\gamma J_{z}\right)  $,
because we are using the eigenbasis of $J_{z}$.

Thus, we can rewrite the Werner state (\ref{WernerState}) in the form
\begin{equation}
\rho_{W}=\frac{1}{4\pi}%
{\displaystyle\int\limits_{0}^{\pi}}
d\theta%
{\displaystyle\int\limits_{0}^{2\pi}}
d\varphi\sin\theta\,[U_{A}\left(  \theta,-\varphi\right)  \otimes U_{B}\left(
\theta,-\varphi\right)  ]\left[  \rho_{A}\otimes\rho_{B}\right]  [U_{A}\left(
\theta,-\varphi\right)  ^{\dagger}\otimes U_{B}\left(  \theta,-\varphi\right)
^{\dagger}], \label{unitarydecomposition}%
\end{equation}
where
\begin{align}
\rho_{A}  &  =\frac{1}{2j+1}{1\!\!1}+\frac{1}{\left(  2j+1\right)  }%
\frac{\left(  2j+1\right)  f-1}{\left(  2j+1\right)  ^{2}-1}%
{\displaystyle\sum\limits_{k=1}^{2j}}
\eta_{k}^{-1}\sqrt{2k+1}T_{0}^{k},\label{digonalA}\\
\rho_{B}  &  =\frac{1}{2j+1}{1\!\!1}+%
{\displaystyle\sum\limits_{k=1}^{2j}}
\eta_{k}\sqrt{2k+1}T_{0}^{k}.
\end{align}
It is easy to see that $\rho_{A}$ and $\rho_{B}$ are diagonal matrices. To
show this \ one uses (\ref{tensoroperators}) and the properties of the
Clebsch-Gordan coefficients, namely $C_{m_{1}qm_{2}}^{jkj}=0$ if $m_{1}+q\neq
m_{2}$. Thus the diagonal elements, i.e. eigenvalues of $\rho_{A}$ and
$\rho_{B}$, are given by%
\begin{equation}
\lambda_{m}\left(  \rho_{A}\right)  =\frac{1}{2j+1}+\frac{1}{\left(
2j+1\right)  ^{3/2}}\frac{\left(  2j+1\right)  f-1}{\left(  2j+1\right)
^{2}-1}%
{\displaystyle\sum\limits_{k=1}^{2j}}
\eta_{k}^{-1}\left(  2k+1\right)  C_{m0m}^{jkj}, \label{eigenvaluesA}%
\end{equation}
and
\begin{equation}
\lambda_{m}\left(  \rho_{B}\right)  =\frac{1}{2j+1}+\frac{1}{\sqrt{2j+1}}%
{\displaystyle\sum\limits_{k=1}^{2j}}
\eta_{k}\left(  2k+1\right)  C_{m0m}^{jkj},\text{ \ }m=j,j-1,...-j.
\label{eigenvaluesB}%
\end{equation}
If all $\lambda_{m}\left(  \rho_{A}\right)  $ and $\lambda_{m}\left(  \rho
_{B}\right)  $ are positive, we found a separable form of the Werner state
(\ref{WernerState}). We note that positivity of $\lambda_{m}\left(  \rho
_{A}\right)  $ and $\lambda_{m}\left(  \rho_{B}\right)  $ for all $m$ implies
that $%
{\displaystyle\sum\limits_{m=-j}^{j}}
\lambda_{m}\left(  \rho_{A}\right)  \lambda_{m}\left(  \rho_{B}\right)
\geq0.$ Owing to the fact that \cite{Biedenharn}%
\begin{equation}%
{\displaystyle\sum\limits_{m=-j}^{j}}
C_{m0m}^{jk_{1}j}C_{m0m}^{jk_{2}j}=\frac{2j+1}{2k+1}\delta_{k_{1}k_{2}} \;\;
\mathrm{and} \; {\displaystyle\sum\limits_{m=-j}^{j}} C_{mqm}^{jkj}=0\;
\mathrm{for}\; k\neq0, \label{columOrthogon}%
\end{equation}
one finds $%
{\displaystyle\sum\limits_{m=-j}^{j}}
\lambda_{m}\left(  \rho_{A}\right)  \lambda_{m}\left(  \rho_{B}\right)  =f.$
Thus $f\geq0$ is a necessary condition for positivity of $\rho_{A}$ and
$\rho_{B}$. It remains to show the sufficient conditions for positivity of
$\rho_{A}$ and $\rho_{B}$. At this point it becomes clear why the asymmetric
form, chosen above for $\rho_{A}$ and $\rho_{B}$, is advantageous: we can
determine the free parameters $\eta_{k}$ from the positivity condition for
$\rho_{B}$ and then find the range of $f$ for which $\rho_{A}$ is positive.

We note that \ for the simplest case of spin $j=1/2$ the present decomposition
is, due to the isomorphism between $SO(3)$ and $SU(2)$, equivalent to
\begin{equation}
\rho_{W}=%
{\displaystyle\int}
dU\left[  U\otimes U\right]  \cdot\left[  \rho_{A}\otimes\rho_{B}\right]
\cdot\left[  U\otimes U\right]  ^{\dagger}, \label{WernerOriginal}%
\end{equation}
where the integral is extended to all unitary operators acting on the
two-dimensional Hilbert space, with $%
{\displaystyle\int}
dU=1$ and $dU$ representing the standard Haar measure on the group $SU\left(
2\right)  $. Hence, we expect that the inequalities $\lambda_{\pm1/2}\left(
\rho_{A,B}\right)  \geq0$ will yield $0\leq f\leq1$. For higher dimensions,
however, there is no such simple argument, and it turns out that for higher
spins the decomposition (\ref{unitarydecomposition}) is a separable
decomposition, i.e. $\rho_{A,B}\geq0,$ only in the range $0\leq f\leq
f_{0}\left(  j\right)  ,$ where $f_{0}\left(  j\right)  <1.$

The presented method can also be used to find a separable forms for more
general states e.g. states which are invariant under product representations
of the group SO(3) of three-dimensional rotations, see \cite{Werner3}.

\section{Examples}

Let us consider explicitly the two lowest dimensions, namely $d=2$ (spin
$j=1/2$) and $d=3$ (spin $j=1$).

\subsection{Case j=1/2}

Here, the index $k$ in Eqs. (\ref{eigenvaluesA}) and (\ref{eigenvaluesB})
takes one value, $k=1$, and we have one free parameter $\eta_{1}$. Thus, for
qubits we have the following inequalities:
\begin{align*}
\lambda_{1/2}\left(  \rho_{B}\right)   &  =\frac{1}{2}\left(  1+\sqrt{6}%
\eta_{1}\right)  \geq0,\\
\lambda_{-1/2}\left(  \rho_{B}\right)   &  =\frac{1}{2}\left(  1-\sqrt{6}%
\eta_{1}\right)  \geq0.
\end{align*}
This leads to the condition
\begin{equation}
\left\vert \eta_{1}\right\vert \leq\frac{1}{\sqrt{6}}. \label{a}%
\end{equation}
Positivity of $\lambda_{\pm1/2}\left(  \rho_{A}\right)  $ translates to the
constraints
\begin{align}
\lambda_{1/2}\left(  \rho_{A}\right)   &  =\frac{1}{2}+\frac{\sqrt{6}}%
{12}\frac{2f-1}{\eta_{1}}\geq0,\label{b}\\
\lambda_{-1/2}\left(  \rho_{A}\right)   &  =\frac{1}{2}-\frac{\sqrt{6}}%
{12}\frac{2f-1}{\eta_{1}}\geq0. \label{c}%
\end{align}
The solution of the inequalities (\ref{a}),(\ref{b}), and (\ref{c}) is
\[
\left\vert \eta_{1}\right\vert \leq\frac{1}{\sqrt{6}}\text{ and }0\leq
f\leq1.
\]
Thus, as mentioned above, for $j=1/2$ our decomposition is valid for the whole
separable range of the Werner state family.

\subsection{Case j=1}

For $j=1$, we have $k=1,2$. We can use two free parameters $\eta_{1},\eta_{2}%
$. Positivity of $\lambda_{m}\left(  \rho_{B}\right)  $ for $m=-1,0,+1$
implies the following inequalities:

\begin{align}
\lambda_{1}\left(  \rho_{B}\right)   &  =\frac{1}{6}\left(  2+3\sqrt{6}%
\eta_{1}+\sqrt{30}\eta_{2}\right)  \geq0\ ,\label{aaasystem}\\
\lambda_{-1}\left(  \rho_{B}\right)   &  =\frac{1}{6}\left(  2-3\sqrt{6}%
\eta_{1}+\sqrt{30}\eta_{2}\right)  \geq0\ ,\nonumber\\
\lambda_{0}\left(  \rho_{B}\right)   &  =\frac{1}{3}\left(  1-\sqrt{30}%
\eta_{2}\right)  \geq0\ .\nonumber
\end{align}

The eigenvalues of $\lambda_{m}\left(  \rho_{A}\right)  $ read
\begin{align}
\lambda_{1}\left(  \rho_{A}\right)   &  =\frac{1}{144}\left(  \frac{\eta
_{1}\left(  \sqrt{30}\left(  3f-1\right)  +48\eta_{2}\right)  +3\sqrt
{6}\left(  3f-1\right)  \eta_{2}}{\eta_{2}\eta_{1}}\right)
\ ,\label{bbbsystem}\\
\lambda_{-1}\left(  \rho_{A}\right)   &  =\frac{1}{144}\left(  \frac{\eta
_{1}\left(  \sqrt{30}\left(  3f-1\right)  +48\eta_{2}\right)  -3\sqrt
{6}\left(  3f-1\right)  \eta_{2}}{\eta_{2}\eta_{1}}\right)  \ ,\nonumber\\
\lambda_{0}\left(  \rho_{A}\right)   &  =\frac{1}{72}\left(  24-\left(
3f-1\right)  \frac{\sqrt{30}}{\eta_{2}}\right)  \ .\nonumber
\end{align}

After a lengthy calculation we conclude that positivity of $\rho_{A}$ holds
for
\[
0\leq f\leq\frac{3}{5}\ ,
\]
i.e. for the case $j=1$ our decomposition is valid only within a certain range
of the parameter $f$, and not for the whole separable interval $0\leq f\leq1$.
Note that our separable decomposition holds for Werner states that are
\textquotedblleft close\textquotedblright\ to entangled states.

\begin{figure}
[ptb]
\begin{center}
\includegraphics[
height=3.8795in,
width=3.7957in
]%
{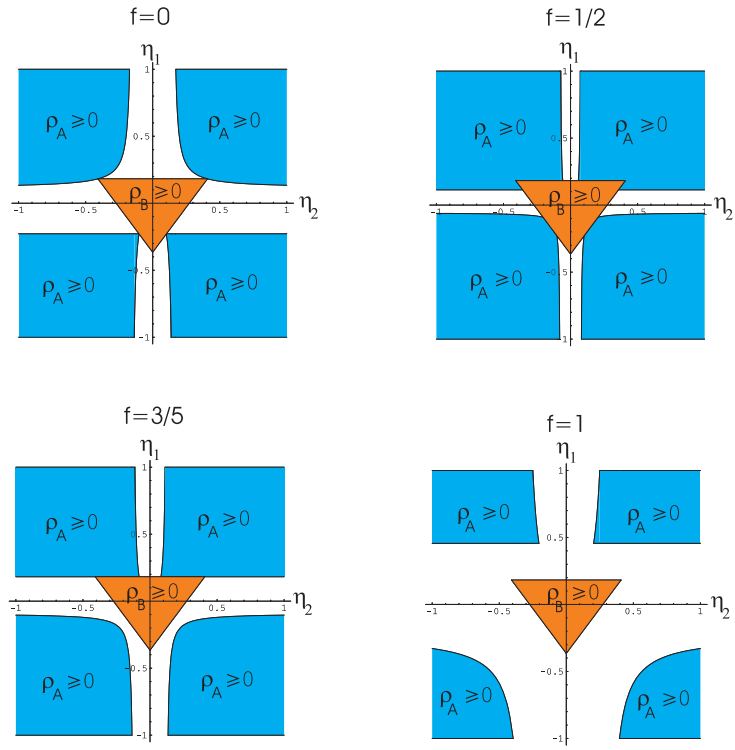}%
\caption{Case spin $j=1$: Solutions of inequalities (\ref{aaasystem}) and
(\ref{bbbsystem}) (filled regions) for different values of $f=0,\frac{1}%
{2},\frac{3}{5}$ and $1.$}%
\label{inequlity}%
\end{center}
\end{figure}

Fig.\ref{inequlity} shows the solutions of inequalities $\lambda_{m}\left(
\rho_{A,B}\right)  \geq0$ for different values of $\ f$ $,$ for spin $j=1$.
One can see that the common area of filled regions vanishes for $f>\frac{3}%
{5}$.%

\section{Conclusions}

In summary, we have used an angular momentum approach to find in principle a
separable decomposition of Werner states in any dimension. Our main idea is to
express the local density operators and the global operator in terms of Wigner
operators. The eigenvalues of the local operators are found to be certain
combinations of Clebsch-Gordan coefficients. Some free parameters allow to
guarantee positivity of the local density operators. For dimension $d=2$ our
decomposition holds for all separable Werner states, while for higher
dimensions it is only valid for a certain range of separable Werner states.
The size of this range depends on the dimension. We verified that our
decomposition is valid for $f=0$ up to dimension $d=5$. It is still an open
task to find a product decomposition of \emph{all} separable Werner states in
any dimension.

We \ would like to thank R. Werner, M. Kleinmann and T. Meyer for informative discussions.


\begin{thebibliography}{9}                                                                                                %


\bibitem {Werner}R.F. Werner, Phys. Rev. A \textbf{40}, 4277 (1989).

\bibitem {db}D. Bru\ss , J. Math. Phys. \textbf{43}, 4237 (2002).

\bibitem {primer}M. Lewenstein et al,
J. Mod. Opt. \textbf{47}, 2841 (2000).

\bibitem {puri}C. H. Bennett, D. P. DiVincenzo, J. A. Smolin, and W. K.
Wootters, Phys. Rev. A \textbf{54}, 3824 (1996).

\bibitem {npt}D. DiVincenzo, P. Shor, J. Smolin, B. Terhal, and A. Thapliyal,
Phys. Rev. A \textbf{61}, 062312 (2000); W. D\"ur, J. I. Cirac, M. Lewenstein,
and D. Bru\ss , Phys. Rev. A \textbf{61}, 062313 (2000).

\bibitem {wootters}W. K. Wootters, Phys. Rev. Lett. \textbf{80}, 2245 (1998).

\bibitem {azuma}H. Azuma and M. Ban, Phys. Rev. A \textbf{73}, 032315 (2006).

\bibitem {Biedenharn}L. C. Biedenharn and J. D. Louck, Angular Momentum in
Quantum Physics, Addison-Wesley, Reading, MA (1981).

\bibitem {Werner3}K. Vollbrecht and R. Werner, Phys. Rev. A \textbf{64},
062307 (2001); H.-P. Breuer, Phys. Rev. A \textbf{71}, 062330 (2005).
\end{thebibliography}
\end{document}